\documentclass[aps,prl,twocolumn,groupedaddress,showpacs,altaffilletter]{revtex4}

\usepackage{graphicx}

\begin{document}

\title{Magnetoelastic-magnetoelectric phase transitions in multiferroic BiFeO$_3$}

\author{S. A. T. Redfern}
\email{satr@esc.cam.ac.uk}
\affiliation{Department of Earth Sciences,
University of Cambridge, Downing Street, Cambridge CB2 3EQ, United Kingdom}
\author{C. Wang}
\altaffiliation[Now at ]{Institute of Physics, Chinese Academy of Sciences, Beijing, 100190 China}\affiliation{Department of Earth Sciences,
University of Cambridge, Downing Street, Cambridge CB2 3EQ, United Kingdom}
\author{J. W. Hong}
\altaffiliation[Also at ]{Department of Engineering Mechanics, Tsinghua University, Beijing,
100084 China}
\affiliation{Department of Earth Sciences,
University of Cambridge, Downing Street, Cambridge CB2 3EQ, United Kingdom}
\author{G. Catalan}
\affiliation{Department of Earth Sciences,
University of Cambridge, Downing Street, Cambridge CB2 3EQ, United Kingdom}
\author{J. F. Scott}
\affiliation{Department of Earth Sciences,
University of Cambridge, Downing Street, Cambridge CB2 3EQ, United Kingdom}

\date{23 June 2008; $Revision: 1.6 $}

\begin{abstract}
Our measured dielectric constant and mechanical response of multiferroic BiFeO$_3$ indicate four phase transitions below room temperature. Features correlate with those reported at 50K (from a peak in the zero-field-cooled magnetic susceptibility) and 230K (from splitting between field-cooled and zero-field-cooled magnetic data~\cite{singh08a}), and those at 140K and 200K (from magnon light scattering cross sections~\cite{singh08b}). The primary order parameter is not the polarization in any of the low-$T$ transitions. Instead, the transition near 200 K shows strong elastic coupling, while that at 50K is fundamentally magnetic, but magnetostrictively coupled to the the lattice. The low-$T$ phase transitions display glassy behaviour. A further anomaly at 140K interpreted as spin reorientation~\cite{singh08b,cazayous07} shows only weakly in dielectric and mechanical studies, indicating that it is predominantly magnetic with little coupling to any of the other order parameters.
\end{abstract}

\pacs{75.80.+q; 77.90.+k; 77.80.Bh;81.40.Jj}

\maketitle


Magnetoelectric multiferroics are materials where ferroelectricity and magnetism coexist and are coupled. These materials are attracting increasing interest due to their possible applications as magnetic detectors and multi-state memory devices, as well as for the rich physics behind the fundamental mechanisms for magnetoelectric coupling. Among the materials being most actively studied, BiFeO$_3$ (ÒBFOÓ) stands out for being the only known room $T$ magnetoelectric multiferroic (ferroelectric, antiferromagnetic), and has some very attractive features such as huge remnant polarization (around 100 $\mu$C/cm$^2$) in both single-crystal~\cite{lebeugle07a,lebeugle07b} and thin-film form~\cite{eerenstein05,wang03} and coupling between the directions of the ferroelectric polarization and the sublattice magnetization~\cite{zhao06}. For this reason, BFO has been touted as a possible active material in the next generation of lead-free ferroelectric materials~\cite{fujitsu06}] and also as the active layer in magnetoelectric memories based on exchange bias~\cite{ramesh07}.

BiFeO$_3$ has a rhombohedrically distorted perovskite structure (space group R3c) at room $T$~\cite{TEAGUE:1970xw} which is both ferroelectric (with polarization along the $\langle$111$\rangle$ directions) and antiferromagnetic. The antiferromagnetic symmetry is itself very exotic, with a long period  (62 nm) cycloidal helix of spins, inconmensurate with the lattice periodicity~\cite{sosnowska82,Zalesskii:2000ak}. The cycloidal rotation cancels almost all macroscopic magnetization, and hence BFO is antiferromagnetic; upon heating, it becomes paramagnetic at $T_N$=640K~\cite{smolensky63}. Around 1100K BFO shows a transition from rhombohedral to orthorhombic symmetry [7] or perhaps monoclinic [8], and around 1200K it becomes cubic. It is not yet certain which of the two phase transitions is the ferroelectric-paraelectric one. On the other hand, there is evidence that the transition to the cubic phase is accompanied by the onset of a metal-insulator transition~\cite{palai08,kornev07}. While the high $T$ phase diagram appears to be quite rich, there has been comparatively little reported for the low-$T$ properties, although it has been suggested~\cite{NAKAMURA:1993rq,Pradhan:2005oq} that spin-glass behaviour occurs amorphous BFO below 200K. However, three recent works~\cite{singh08a,singh08b,cazayous07} indicate further complexity in the low-$T$ nature of BFO. 

The first of these~\cite{singh08a} reports the magnetic properties of BFO single crystals at low temperatures. It was found that the field-cooled and zero-field cooled magnetization became different at temperatures below $\approx$ 250K, as can subsequently be seen in earlier data~\cite{Naganuma:2007om,Siwach:2007xu}. Furthermore, there is a distinct peak in the zero-field-cooled magnetization around 50K. Below this lower $T$, the temperature-frequency dependence of magnetic susceptibility indicates a possible spin glass transition, although the authors noticed that the behavior was unusual in that susceptibility increases with increasing frequency. It was also reported that the critical exponent describing the slowing down of the glassy dynamics ($z\nu$), was very small ($\approx$ 1.4), much closer to the value expected in a mean-field system (where  $z\nu$=2) than in a classic Ising-type magnetic spin glass  ($z\nu$=7-10)~\cite{KIRKPATRICK:1978pt}.  On this basis, it was suggested that the spin glass transition may be coupled to a long-range order parameter responsible for its mean-field behaviour. 

Later, two contemporaneous and independent Raman studies reported electromagnon Raman spectra in BFO~\cite{singh08b,cazayous07}.  Singh et al.~\cite{singh08b} noted two anomalies in the Raman frequency, intensity, and linewidth: one strong at $\approx$200K, one very weak at $\approx$140K. Cazayous et al. reported a strong anomaly at 140K, and only a keen eye might discern a very small anomaly at 200K. It is interesting that the strength of the anomalies was so different. However, in more recent work they found both transitions with equal magnon cross section divergences by changing their scattering geometry, showing that the differences are not sample-dependent but orientation-dependent. Two separate spin reorientation transition temperatures might be expected since the spins begin to rotate out of a plane at the upper one, while at the lower one the spins have rotated fully 90$^\circ$, so that they are orthogonal to the original plane. Singh et al.'s~\cite{singh08a} data are compatible (in respect of both peak intensity divergence and linewidth narrowing) with the original study of spin fluctuations in uniaxial antiferromagnets by Schulhof et al.~\cite{SCHULHOF:1970kh}.  They found that fluctuations along the uniaxial direction diverge and exhibit critical slowing down (spectral narrowing) approaching $T_c$, with critical exponents $\nu$ = 0.63 and $\gamma$ (susceptibility) = 1.24 for the longitudinal fluctuations and $\nu$ = 0.63 and $\gamma$ = 1.47 for the transverse fluctuations. Note that the transverse fluctuations are not truly divergent but extrapolate to a divergence $\approx$2K below $T_c$.


\begin{figure}
\includegraphics{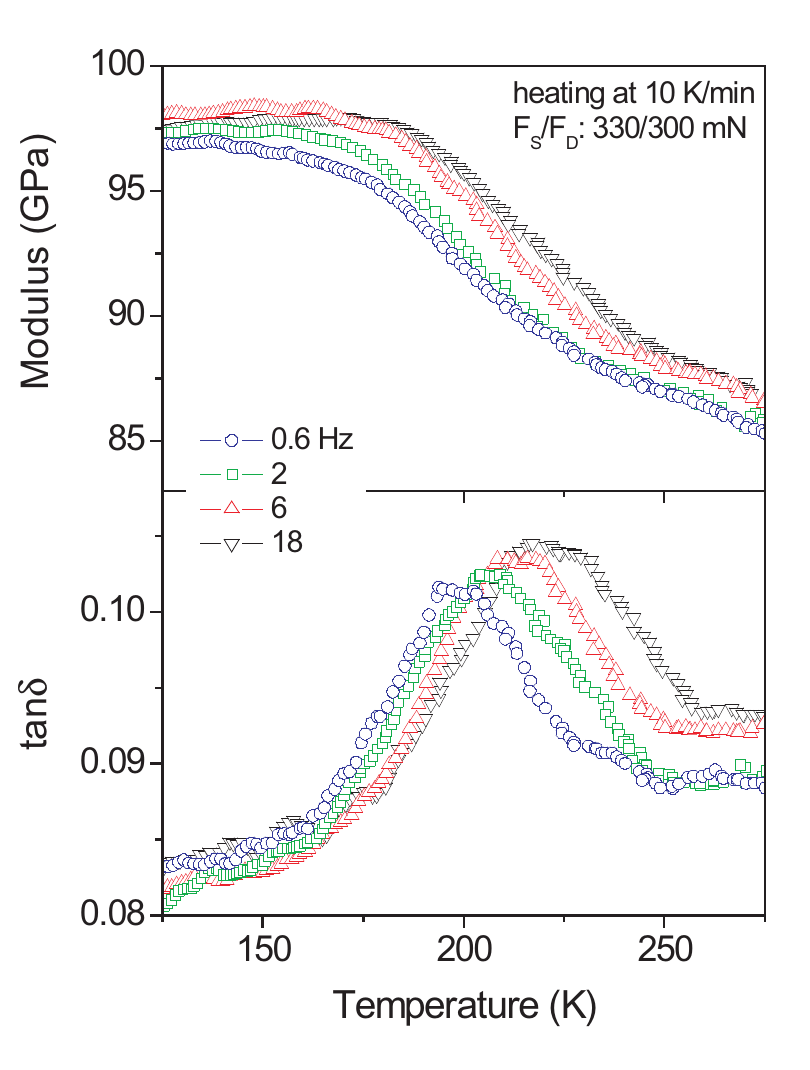}
\caption{\label{DMA_freq} Frequency dependence of the dynamic mechanical modulus and tan$\delta$ of BiFeO$_3$ as a function of $T$. The large peak in tan$\delta$ and concomitant relaxation in modulus on heating through 200K shows significant frequency-temperature dispersion.}
\end{figure}

In order to understand further the nature of these low-$T$ anomalies, we have studied them from the point of view of the two other order parameters of this multiferroic material: ferroelasticity and ferroelectricity. Dynamical Mechanical Analysis of ceramic BFO bars was performed in a three-point bending configuration using samples that were 2mm wide, 0.5mm thick, and 5mm between knife edges. Force between  30 and 630 mN were applied dynamically, generating strains of the order of 10$^{-4}$, at frequencies between 0.6 and 18 Hz and $T$s between 120 and 290K (Fig.~\ref{DMA_freq}). The YoungÕs modulus and loss angle (indicative of a mechanical dissipative process, such as domain wall friction~\cite{Wang:2006sw}) show stiffening of $\approx$10 $\%$ upon cooling accompanied by a simultaneous peak in dissipation, tan$\delta$, indicative of transition from a dynamically relaxed system above 230K to an unrelaxed state below this temperature. There is strong frequency dependence of the mechanical relaxation $T$, signaling either a true glassy transition or else a defect-related relaxation. The limited range of frequencies accessible in our measurement makes it difficult to compare Vogel-Fulcher (glassy freezing) and Arrhenius (thermal activation of defect states) models for the response. None the less, analysis of the frequency dependence of the peak in tan$\delta$ around 200K is consistent with an activation energy of 0.59(9) eV which is smaller than that expected for mechanical relaxation due to oxygen vacancies in perovskite oxides (typically around 0.7 to 1.1 eV~\cite{Harrison:2004et}).

\begin{figure}

\includegraphics[width=0.42\textwidth]{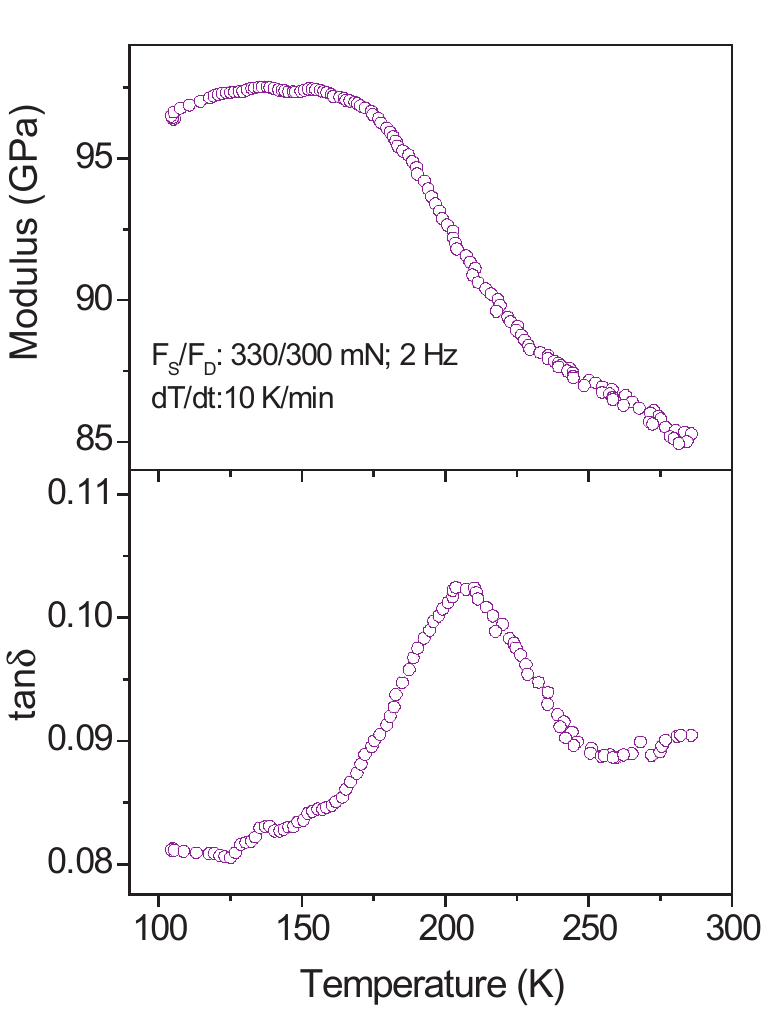}
\caption{\label{DMA_2Hz} The mechanical modulus and tan$\delta$ of BiFeO$_3$ measured at 2Hz. A small anomaly in tan$\delta$ can be seen at 140K.}
\end{figure}

The DMA data also show a very weak anomaly, seen most clearly in the data for  tan$\delta$ collected at 2Hz (Fig.~\ref{DMA_2Hz}), at 140K. The weakness of the 140K feature and strength of the one around 200K contrasts markedly with the relative scale of the features seen at these $T$s in the electromagnon spectra, which may indicate variable coupling between strain and the primary order parameter at the lower-$T$ anomaly.

\begin{figure}

\includegraphics[width=0.45\textwidth]{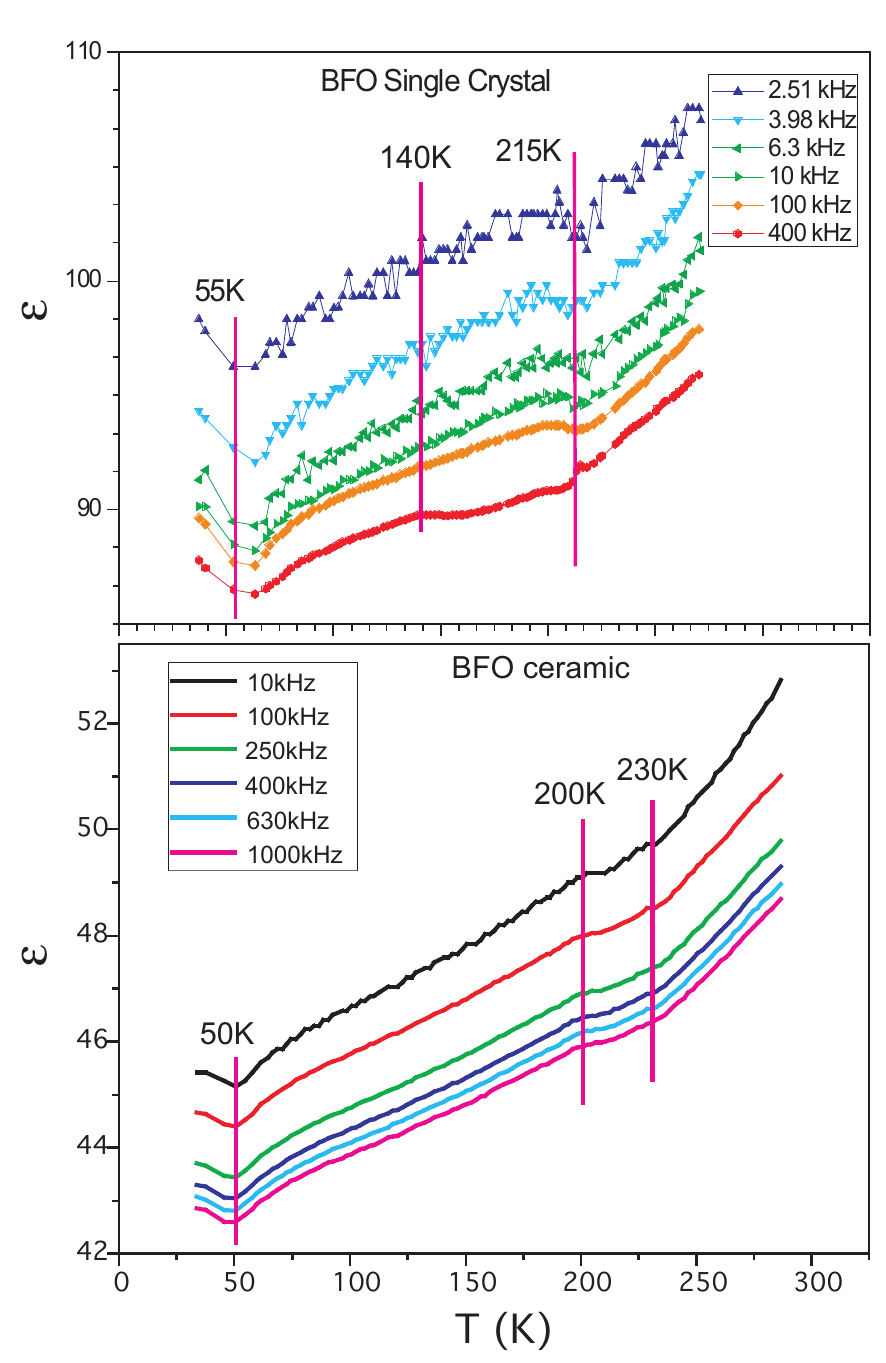}
\caption{\label{Dielectric} Dielectric constant of BiFeO$_3$ at low-$T$. A small anomaly at 140K can be seen in the single crystal data at the highest frequency.}
\end{figure}

The dielectric constant was also measured as a function of $T$ and frequency for single crystals and for ceramics (see Fig.~\ref{Dielectric}). The single crystals were cut with faces perpendicular to \{110\} pseudocubic planes. The dielectric constant shows subtle but unambiguous anomalies around 50K, 200 and 230K. The fact that they are subtle may explain why they may have gone unnoticed in spite of extensive research on this compound in recent years. It also indicates that the primary order parameter in these transitions is almost certainly not the polarization, and that they reflect indirect coupling to the primary order parameter. While a single broad peak, strongly dependent on frequency, is seen around 200K in DMA, two resolved anomalies (independent of frequency) are seen in the dielectirc data. 

\begin{figure}

\includegraphics[width=0.50\textwidth]{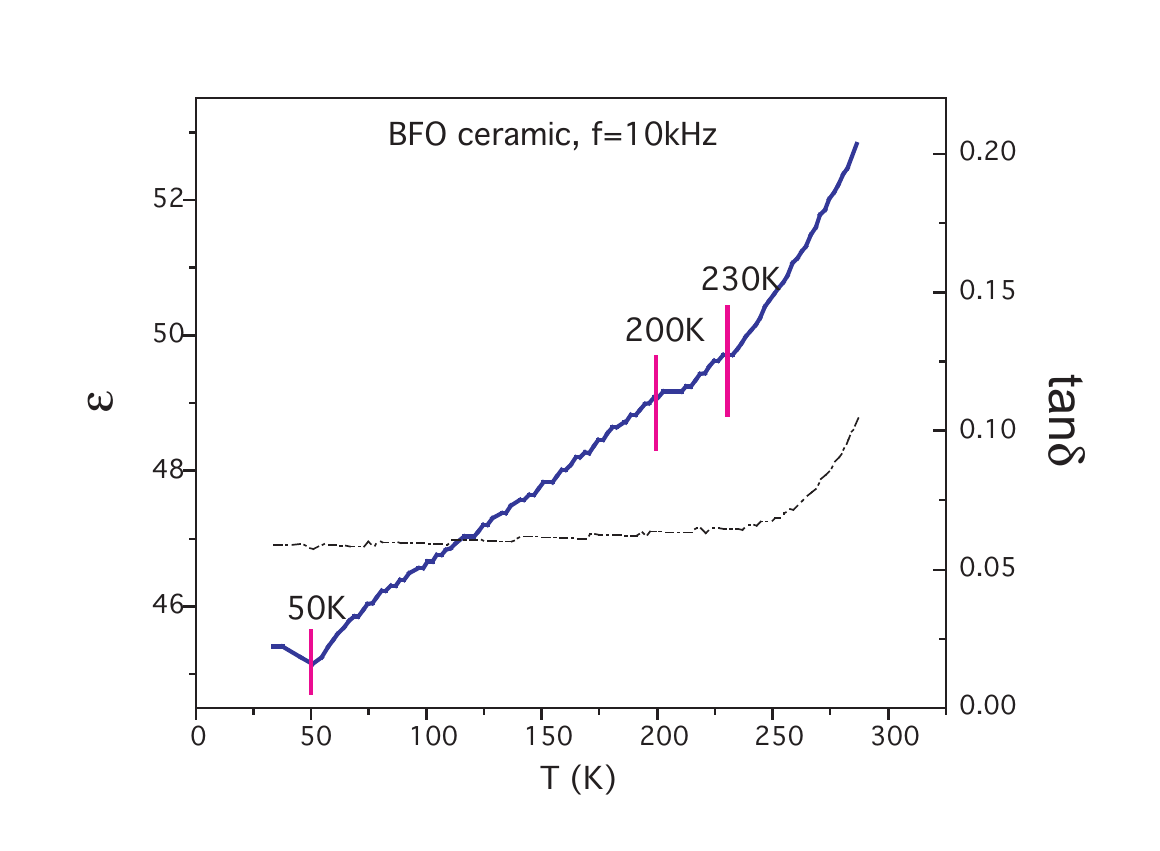}
\caption{\label{TanD} Dielectric constant and loss of BiFeO$_3$ at low-$T$, measured at 10kHz.}
\end{figure}

We note that there is no strong evidence of the anomaly at 140K except at the highest frequency used (400 kHz).  This suggests that further studies in the MHz regime should be carried out on single crystal samples; it may be that this transition is insensitive to low frequency probes. The dielectric constant (capacitance) shows some low-frequency dispersion, probably due to space-charge, but nevertheless the position of the anomalies appears to be relatively frequency-independent, contrary to the glassiness observed in the magnetic and mechanical data. It is also worth mentioning that although the space charge increases the dispersion and the dielelectric losses, these remain low (tan$\delta$\textless0.1) (Fig.~\ref{TanD}) and furthermore (perhaps surprisingly) there is no obvious sign of change in the losses at the critical temperatures, ruling out a conductive artifact~\cite{Catalan:2006le}.

Both the mechanical and dielectric data show only very weak evidence of the transition at 140K, which suggests that this transition is predominantly magnetic. The transitions above $\approx$200K, on the other hand, have now been observed in different samples (single crystals and ceramics) and with five different techniques (magnetization, susceptibility, Raman, impedance and mechanical spectroscopy). If either is due to defects, they are of a very pervasive but undetectable nature:  XRD studies did not detect any impurities or second-phases in any of our samples. Furthermore, the mechanical response seems very large to be caused by to defects alone, suggesting instead that the primary order parameter may be a structural one, coupled weakly to the polarization (hence the dielectric anomaly) and to the magnetization (hence the electromagnon and magnetization anomalies). In the dielectric data two small discrete anomalies at 200 and 230K are seen while the mechanical relaxation at this temperature is broad and large.

The low-$T$ anomaly indielectric response at 50K is weak but nonetheless clear, showing that  the magnetic order parameter is coupled to the polar one. Since the latter is a long-range order parameter, this coupling may account for the apparent (and very unusual) mean-field nature of the low-$T$ spin glass transition.  We note also that if this material is indeed a spin-glass, it would be acentric.  Fischer and Hertz~\cite{fischer91} have emphasized that no published theories apply to spin glasses lacking an inversion center, and further, that such glasses cannot possibly be Ising-like.

Our results indicate that BiFeO$_3$ is remarkably similar to the orthoferrite ErFeO$_3$, which has $T_N$ = 633K and spin-reorientation transitions at 90K and 103K [compared with $T_N$ = 643K and spin reorientation transitions at 140K and 200K in BiFeO$_3$.  According to the theory of Koshizuka and Ushioda~\cite{KOSHIZUKA:1980sj}  one might expect the magnon frequency $f$ to drop to zero at $T_{reorientation}$ and the magnon integrated cross section to diverge due to enhanced thermal population k$T$/h$f$; however Koshizuka and Ushioda report  [PRB 22, 5394 (1980)] a smaller (50\%) magnon frequency decrease and consequently a cross section increase of only about $\times 5$, and attribute this to phonon-magnon coupling (magnetostriction).  Singh et al.'s~\cite{singh08b} magnon frequency decrease is only $\approx$5 \% and the peak scattering amplitude increases by a maximum of $\times$5 to $\times$10  - but the linewidth decreases by at least $\times$5 (resolution limited). Thus the integrated cross section changes by only a small amount. This is compatible with the strong mechanical losses we see around 200K (and somewhat weaker at 140K) and implies very large striction. This is compatible with respect to linewidth narrowing and peak amplitude divergence with the spin-fluctuation models of Koshizuka and Ushioda and Schulhof et al. ~\cite{KOSHIZUKA:1980sj,SCHULHOF:1970kh}; however, the  linewidth narrowing and peak amplitude divergence imply critical fluctuations not explicit in their original papers.

In summary, we show that the anomalies in bismuth ferrite at cryogenic $T$s can be interpreted as phase transitions at 50K (magnetic, but glassy and with magnetoelectric coupling), 140K (dominantly magnetic), 200K (magnetoelastic), and 230K (magnetic; glassy).  The coupling between magnetic, electric and elastic order parameters varies according to the length scale of the transition dynamics in each case: dipolar and strain coupling leads to the possibility of long-range correlations dominating the spin glass behavior, resulting in a rare example of an acentric mean-field spin glass in BFO. It would appear that a new model for non-Ising spin glasses must now be developed if the low-$T$ magnetoelectric properties of BiFeO$_3$ are to be adequately explained.

\begin{acknowledgments}
We are grateful to R Palai and H Schmid who provided ceramic and single crystal samples of BFO.
\end{acknowledgments}

\bibliography{mag_elas_elec_BFO}

\begin{thebibliography}{27}
\expandafter\ifx\csname natexlab\endcsname\relax\def\natexlab#1{#1}\fi
\expandafter\ifx\csname bibnamefont\endcsname\relax
  \def\bibnamefont#1{#1}\fi
\expandafter\ifx\csname bibfnamefont\endcsname\relax
  \def\bibfnamefont#1{#1}\fi
\expandafter\ifx\csname citenamefont\endcsname\relax
  \def\citenamefont#1{#1}\fi
\expandafter\ifx\csname url\endcsname\relax
  \def\url#1{\texttt{#1}}\fi
\expandafter\ifx\csname urlprefix\endcsname\relax\def\urlprefix{URL }\fi
\providecommand{\bibinfo}[2]{#2}
\providecommand{\eprint}[2][]{\url{#2}}

\bibitem[{\citenamefont{Singh et~al.}(2008{\natexlab{a}})\citenamefont{Singh,
  Prellier, Singh, Katiyar, and Scott}}]{singh08a}
\bibinfo{author}{\bibfnamefont{M.~K.} \bibnamefont{Singh}},
  \bibinfo{author}{\bibfnamefont{W.}~\bibnamefont{Prellier}},
  \bibinfo{author}{\bibfnamefont{M.~P.} \bibnamefont{Singh}},
  \bibinfo{author}{\bibfnamefont{R.~S.} \bibnamefont{Katiyar}},
  \bibnamefont{and} \bibinfo{author}{\bibfnamefont{J.~F.} \bibnamefont{Scott}},
  \bibinfo{journal}{Physical Review B} \textbf{\bibinfo{volume}{77}},
  \bibinfo{pages}{144403} (\bibinfo{year}{2008}{\natexlab{a}}).

\bibitem[{\citenamefont{Singh et~al.}(2008{\natexlab{b}})\citenamefont{Singh,
  Katiyar, and Scott}}]{singh08b}
\bibinfo{author}{\bibfnamefont{M.~K.} \bibnamefont{Singh}},
  \bibinfo{author}{\bibfnamefont{R.~S.} \bibnamefont{Katiyar}},
  \bibnamefont{and} \bibinfo{author}{\bibfnamefont{J.~F.} \bibnamefont{Scott}},
  \bibinfo{journal}{Journal of Physics-Condensed Matter}
  \textbf{\bibinfo{volume}{20}}, \bibinfo{pages}{25203}
  (\bibinfo{year}{2008}{\natexlab{b}}).

\bibitem[{\citenamefont{Cazayous et~al.}(2007)\citenamefont{Cazayous, Gallais,
  Sacuto, {de Sousa}, Lebeugle, and Colson}}]{cazayous07}
\bibinfo{author}{\bibfnamefont{M.}~\bibnamefont{Cazayous}},
  \bibinfo{author}{\bibfnamefont{Y.}~\bibnamefont{Gallais}},
  \bibinfo{author}{\bibfnamefont{A.}~\bibnamefont{Sacuto}},
  \bibinfo{author}{\bibfnamefont{R.}~\bibnamefont{{de Sousa}}},
  \bibinfo{author}{\bibfnamefont{D.}~\bibnamefont{Lebeugle}}, \bibnamefont{and}
  \bibinfo{author}{\bibfnamefont{D.}~\bibnamefont{Colson}}
  (\bibinfo{year}{2007}),
  \urlprefix\url{http://www.citebase.org/abstract?id=oai:arXiv.org:0712.3044}.

\bibitem[{\citenamefont{Lebeugle
  et~al.}(2007{\natexlab{a}})\citenamefont{Lebeugle, Colson, Forget, and
  Viret}}]{lebeugle07a}
\bibinfo{author}{\bibfnamefont{D.}~\bibnamefont{Lebeugle}},
  \bibinfo{author}{\bibfnamefont{D.}~\bibnamefont{Colson}},
  \bibinfo{author}{\bibfnamefont{A.}~\bibnamefont{Forget}}, \bibnamefont{and}
  \bibinfo{author}{\bibfnamefont{M.}~\bibnamefont{Viret}},
  \bibinfo{journal}{Applied Physics Letters} \textbf{\bibinfo{volume}{91}},
  \bibinfo{pages}{022907} (\bibinfo{year}{2007}{\natexlab{a}}).

\bibitem[{\citenamefont{Lebeugle
  et~al.}(2007{\natexlab{b}})\citenamefont{Lebeugle, Colson, Forget, Viret,
  Bonville, Marucco, and Fusil}}]{lebeugle07b}
\bibinfo{author}{\bibfnamefont{D.}~\bibnamefont{Lebeugle}},
  \bibinfo{author}{\bibfnamefont{D.}~\bibnamefont{Colson}},
  \bibinfo{author}{\bibfnamefont{A.}~\bibnamefont{Forget}},
  \bibinfo{author}{\bibfnamefont{M.}~\bibnamefont{Viret}},
  \bibinfo{author}{\bibfnamefont{P.}~\bibnamefont{Bonville}},
  \bibinfo{author}{\bibfnamefont{J.~F.} \bibnamefont{Marucco}},
  \bibnamefont{and} \bibinfo{author}{\bibfnamefont{S.}~\bibnamefont{Fusil}},
  \bibinfo{journal}{Physical Review B} \textbf{\bibinfo{volume}{76}},
  \bibinfo{pages}{024116} (\bibinfo{year}{2007}{\natexlab{b}}).

\bibitem[{\citenamefont{Eerenstein et~al.}(2005)\citenamefont{Eerenstein,
  Morrison, Dho, Blamire, Scott, and Mathur}}]{eerenstein05}
\bibinfo{author}{\bibfnamefont{W.}~\bibnamefont{Eerenstein}},
  \bibinfo{author}{\bibfnamefont{F.~D.} \bibnamefont{Morrison}},
  \bibinfo{author}{\bibfnamefont{J.}~\bibnamefont{Dho}},
  \bibinfo{author}{\bibfnamefont{M.~G.} \bibnamefont{Blamire}},
  \bibinfo{author}{\bibfnamefont{J.~F.} \bibnamefont{Scott}}, \bibnamefont{and}
  \bibinfo{author}{\bibfnamefont{N.~D.} \bibnamefont{Mathur}},
  \bibinfo{journal}{Science} \textbf{\bibinfo{volume}{307}},
  \bibinfo{pages}{1203} (\bibinfo{year}{2005}).

\bibitem[{\citenamefont{Wang et~al.}(2003)\citenamefont{Wang, Neaton, Zheng,
  Nagarajan, Ogale, Liu, Viehland, Vaithyanathan, Schlom, Waghmare
  et~al.}}]{wang03}
\bibinfo{author}{\bibfnamefont{J.}~\bibnamefont{Wang}},
  \bibinfo{author}{\bibfnamefont{J.~B.} \bibnamefont{Neaton}},
  \bibinfo{author}{\bibfnamefont{H.}~\bibnamefont{Zheng}},
  \bibinfo{author}{\bibfnamefont{V.}~\bibnamefont{Nagarajan}},
  \bibinfo{author}{\bibfnamefont{S.~B.} \bibnamefont{Ogale}},
  \bibinfo{author}{\bibfnamefont{B.}~\bibnamefont{Liu}},
  \bibinfo{author}{\bibfnamefont{D.}~\bibnamefont{Viehland}},
  \bibinfo{author}{\bibfnamefont{V.}~\bibnamefont{Vaithyanathan}},
  \bibinfo{author}{\bibfnamefont{D.~G.} \bibnamefont{Schlom}},
  \bibinfo{author}{\bibfnamefont{U.~V.} \bibnamefont{Waghmare}},
  \bibnamefont{et~al.}, \bibinfo{journal}{Science}
  \textbf{\bibinfo{volume}{299}}, \bibinfo{pages}{1719} (\bibinfo{year}{2003}).

\bibitem[{\citenamefont{Zhao et~al.}(2006)\citenamefont{Zhao, Scholl,
  Zavaliche, Lee, Barry, Doran, Cruz, Chu, Ederer, Spaldin et~al.}}]{zhao06}
\bibinfo{author}{\bibfnamefont{T.}~\bibnamefont{Zhao}},
  \bibinfo{author}{\bibfnamefont{A.}~\bibnamefont{Scholl}},
  \bibinfo{author}{\bibfnamefont{F.}~\bibnamefont{Zavaliche}},
  \bibinfo{author}{\bibfnamefont{K.}~\bibnamefont{Lee}},
  \bibinfo{author}{\bibfnamefont{M.}~\bibnamefont{Barry}},
  \bibinfo{author}{\bibfnamefont{A.}~\bibnamefont{Doran}},
  \bibinfo{author}{\bibfnamefont{M.~P.} \bibnamefont{Cruz}},
  \bibinfo{author}{\bibfnamefont{Y.~H.} \bibnamefont{Chu}},
  \bibinfo{author}{\bibfnamefont{C.}~\bibnamefont{Ederer}},
  \bibinfo{author}{\bibfnamefont{N.~A.} \bibnamefont{Spaldin}},
  \bibnamefont{et~al.}, \bibinfo{journal}{Nature Materials}
  \textbf{\bibinfo{volume}{5}}, \bibinfo{pages}{823} (\bibinfo{year}{2006}).

\bibitem[{fuj()}]{fujitsu06}
\bibinfo{note}{Fujitsu press release 2 August 2006 (Sunnyvale CA):"Fujitsu and
  Tokyo Institute of Technology Announce the Development of New Material for
  256 Mbit FeRAM Using 65-nanometer Technology"}.

\bibitem[{ram()}]{ramesh07}
\bibinfo{note}{R. Ramesh "Whither Oxide Electronics?" Turnbull Lecture, MRS
  (Boston) 2007}.

\bibitem[{\citenamefont{Teague et~al.}(1970)\citenamefont{Teague, Gerson, and
  James}}]{TEAGUE:1970xw}
\bibinfo{author}{\bibfnamefont{J.}~\bibnamefont{Teague}},
  \bibinfo{author}{\bibfnamefont{R.}~\bibnamefont{Gerson}}, \bibnamefont{and}
  \bibinfo{author}{\bibfnamefont{W.}~\bibnamefont{James}},
  \bibinfo{journal}{Solid State Communications} \textbf{\bibinfo{volume}{8}},
  \bibinfo{pages}{1073} (\bibinfo{year}{1970}).

\bibitem[{\citenamefont{Sosnowska et~al.}(1982)\citenamefont{Sosnowska,
  Peterlinneumaier, and Steichele}}]{sosnowska82}
\bibinfo{author}{\bibfnamefont{I.}~\bibnamefont{Sosnowska}},
  \bibinfo{author}{\bibfnamefont{T.}~\bibnamefont{Peterlinneumaier}},
  \bibnamefont{and}
  \bibinfo{author}{\bibfnamefont{E.}~\bibnamefont{Steichele}},
  \bibinfo{journal}{Journal of Physics C-Solid State Physics}
  \textbf{\bibinfo{volume}{15}}, \bibinfo{pages}{4835} (\bibinfo{year}{1982}).

\bibitem[{\citenamefont{Zalesskii et~al.}(2000)\citenamefont{Zalesskii,
  Zvezdin, Frolov, and Bush}}]{Zalesskii:2000ak}
\bibinfo{author}{\bibfnamefont{A.}~\bibnamefont{Zalesskii}},
  \bibinfo{author}{\bibfnamefont{A.}~\bibnamefont{Zvezdin}},
  \bibinfo{author}{\bibfnamefont{A.}~\bibnamefont{Frolov}}, \bibnamefont{and}
  \bibinfo{author}{\bibfnamefont{A.}~\bibnamefont{Bush}},
  \bibinfo{journal}{JETP Letters} \textbf{\bibinfo{volume}{71}},
  \bibinfo{pages}{465} (\bibinfo{year}{2000}).

\bibitem[{\citenamefont{Smolenski and Yudin}(1963)}]{smolensky63}
\bibinfo{author}{\bibfnamefont{G.}~\bibnamefont{Smolenski}} \bibnamefont{and}
  \bibinfo{author}{\bibfnamefont{V.}~\bibnamefont{Yudin}},
  \bibinfo{journal}{Sov. Phys. JETP} \textbf{\bibinfo{volume}{16}},
  \bibinfo{pages}{622} (\bibinfo{year}{1963}).

\bibitem[{\citenamefont{Palai et~al.}(2008)\citenamefont{Palai, Katiyar,
  Schmid, Tissot, Clark, Robertson, Redfern, Catalan, and Scott}}]{palai08}
\bibinfo{author}{\bibfnamefont{R.}~\bibnamefont{Palai}},
  \bibinfo{author}{\bibfnamefont{R.~S.} \bibnamefont{Katiyar}},
  \bibinfo{author}{\bibfnamefont{H.}~\bibnamefont{Schmid}},
  \bibinfo{author}{\bibfnamefont{P.}~\bibnamefont{Tissot}},
  \bibinfo{author}{\bibfnamefont{S.~J.} \bibnamefont{Clark}},
  \bibinfo{author}{\bibfnamefont{J.}~\bibnamefont{Robertson}},
  \bibinfo{author}{\bibfnamefont{S.~A.~T.} \bibnamefont{Redfern}},
  \bibinfo{author}{\bibfnamefont{G.}~\bibnamefont{Catalan}}, \bibnamefont{and}
  \bibinfo{author}{\bibfnamefont{J.~F.} \bibnamefont{Scott}},
  \bibinfo{journal}{Physical Review B} \textbf{\bibinfo{volume}{77}}
  (\bibinfo{year}{2008}).

\bibitem[{\citenamefont{Kornev et~al.}(2007)\citenamefont{Kornev, Lisenkov,
  Haumont, Dkhil, and Bellaiche}}]{kornev07}
\bibinfo{author}{\bibfnamefont{I.~A.} \bibnamefont{Kornev}},
  \bibinfo{author}{\bibfnamefont{S.}~\bibnamefont{Lisenkov}},
  \bibinfo{author}{\bibfnamefont{R.}~\bibnamefont{Haumont}},
  \bibinfo{author}{\bibfnamefont{B.}~\bibnamefont{Dkhil}}, \bibnamefont{and}
  \bibinfo{author}{\bibfnamefont{L.}~\bibnamefont{Bellaiche}},
  \bibinfo{journal}{Physics Review Letters} \textbf{\bibinfo{volume}{99}},
  \bibinfo{pages}{227602} (\bibinfo{year}{2007}).

\bibitem[{\citenamefont{Nakamura et~al.}(1993)\citenamefont{Nakamura, Soeya,
  Ikeda, and Tanaka}}]{NAKAMURA:1993rq}
\bibinfo{author}{\bibfnamefont{S.}~\bibnamefont{Nakamura}},
  \bibinfo{author}{\bibfnamefont{S.}~\bibnamefont{Soeya}},
  \bibinfo{author}{\bibfnamefont{N.}~\bibnamefont{Ikeda}}, \bibnamefont{and}
  \bibinfo{author}{\bibfnamefont{M.}~\bibnamefont{Tanaka}},
  \bibinfo{journal}{Journal of Applied Physics} \textbf{\bibinfo{volume}{74}},
  \bibinfo{pages}{5652} (\bibinfo{year}{1993}).

\bibitem[{\citenamefont{Pradhan et~al.}(2005)\citenamefont{Pradhan, Zhang,
  Hunter, Dadson, Loiutts, Katiyar, Zhang, Sellmyer, Roy, Cui
  et~al.}}]{Pradhan:2005oq}
\bibinfo{author}{\bibfnamefont{A.}~\bibnamefont{Pradhan}},
  \bibinfo{author}{\bibfnamefont{K.}~\bibnamefont{Zhang}},
  \bibinfo{author}{\bibfnamefont{D.}~\bibnamefont{Hunter}},
  \bibinfo{author}{\bibfnamefont{J.}~\bibnamefont{Dadson}},
  \bibinfo{author}{\bibfnamefont{P.}~\bibnamefont{Loiutts}, \bibfnamefont{G.B
  .and~Bhattacharya}},
  \bibinfo{author}{\bibfnamefont{R.}~\bibnamefont{Katiyar}},
  \bibinfo{author}{\bibfnamefont{J.}~\bibnamefont{Zhang}},
  \bibinfo{author}{\bibfnamefont{D.}~\bibnamefont{Sellmyer}},
  \bibinfo{author}{\bibfnamefont{U.}~\bibnamefont{Roy}},
  \bibinfo{author}{\bibfnamefont{Y.}~\bibnamefont{Cui}}, \bibnamefont{et~al.},
  \bibinfo{journal}{Journal of Applied Physics} \textbf{\bibinfo{volume}{97}}
  (\bibinfo{year}{2005}).

\bibitem[{\citenamefont{Naganuma and Okamura}(2007)}]{Naganuma:2007om}
\bibinfo{author}{\bibfnamefont{H.}~\bibnamefont{Naganuma}} \bibnamefont{and}
  \bibinfo{author}{\bibfnamefont{S.}~\bibnamefont{Okamura}},
  \bibinfo{journal}{Journal of Applied Physics} \textbf{\bibinfo{volume}{101}}
  (\bibinfo{year}{2007}).

\bibitem[{\citenamefont{Siwach et~al.}(2007)\citenamefont{Siwach, Singh, Singh,
  and Srivastava}}]{Siwach:2007xu}
\bibinfo{author}{\bibfnamefont{P.~K.} \bibnamefont{Siwach}},
  \bibinfo{author}{\bibfnamefont{H.~K.} \bibnamefont{Singh}},
  \bibinfo{author}{\bibfnamefont{J.}~\bibnamefont{Singh}}, \bibnamefont{and}
  \bibinfo{author}{\bibfnamefont{O.~N.} \bibnamefont{Srivastava}},
  \bibinfo{journal}{Applied Physics Letters} \textbf{\bibinfo{volume}{91}}
  (\bibinfo{year}{2007}).

\bibitem[{\citenamefont{Kirkpatrick and
  Sherrington}(1978)}]{KIRKPATRICK:1978pt}
\bibinfo{author}{\bibfnamefont{S.}~\bibnamefont{Kirkpatrick}} \bibnamefont{and}
  \bibinfo{author}{\bibfnamefont{D.}~\bibnamefont{Sherrington}},
  \bibinfo{journal}{Physical Review B} \textbf{\bibinfo{volume}{17}},
  \bibinfo{pages}{4384} (\bibinfo{year}{1978}).

\bibitem[{\citenamefont{Schulhof et~al.}(1970)\citenamefont{Schulhof, Heller,
  Nathans, and Linz}}]{SCHULHOF:1970kh}
\bibinfo{author}{\bibfnamefont{M.}~\bibnamefont{Schulhof}},
  \bibinfo{author}{\bibfnamefont{P.}~\bibnamefont{Heller}},
  \bibinfo{author}{\bibfnamefont{R.}~\bibnamefont{Nathans}}, \bibnamefont{and}
  \bibinfo{author}{\bibfnamefont{A.}~\bibnamefont{Linz}},
  \bibinfo{journal}{Physical Review B} \textbf{\bibinfo{volume}{1}},
  \bibinfo{pages}{2304} (\bibinfo{year}{1970}).

\bibitem[{\citenamefont{Wang et~al.}(2006)\citenamefont{Wang, Redfern,
  Daraktchiev, and Harrison}}]{Wang:2006sw}
\bibinfo{author}{\bibfnamefont{C.}~\bibnamefont{Wang}},
  \bibinfo{author}{\bibfnamefont{S.~A.~T.} \bibnamefont{Redfern}},
  \bibinfo{author}{\bibfnamefont{M.}~\bibnamefont{Daraktchiev}},
  \bibnamefont{and} \bibinfo{author}{\bibfnamefont{R.~J.}
  \bibnamefont{Harrison}}, \bibinfo{journal}{Applied Physics Letters}
  \textbf{\bibinfo{volume}{89}} (\bibinfo{year}{2006}).

\bibitem[{\citenamefont{Harrison et~al.}(2004)\citenamefont{Harrison, Redfern,
  Buckley, and Salje}}]{Harrison:2004et}
\bibinfo{author}{\bibfnamefont{R.}~\bibnamefont{Harrison}},
  \bibinfo{author}{\bibfnamefont{S.}~\bibnamefont{Redfern}},
  \bibinfo{author}{\bibfnamefont{A.}~\bibnamefont{Buckley}}, \bibnamefont{and}
  \bibinfo{author}{\bibfnamefont{E.}~\bibnamefont{Salje}},
  \bibinfo{journal}{Journal of Applied Physics} \textbf{\bibinfo{volume}{95}},
  \bibinfo{pages}{1706} (\bibinfo{year}{2004}).

\bibitem[{\citenamefont{Catalan}(2006)}]{Catalan:2006le}
\bibinfo{author}{\bibfnamefont{G.}~\bibnamefont{Catalan}},
  \bibinfo{journal}{Applied Physics Letters} \textbf{\bibinfo{volume}{88}}
  (\bibinfo{year}{2006}).

\bibitem[{\citenamefont{H.Fischer and Hertz}(1991)}]{fischer91}
\bibinfo{author}{\bibfnamefont{K.}~\bibnamefont{H.Fischer}} \bibnamefont{and}
  \bibinfo{author}{\bibfnamefont{J.}~\bibnamefont{Hertz}},
  \emph{\bibinfo{title}{Spin Glasses}}, Cambridge Studies in Magnetism
  (\bibinfo{publisher}{Cambridge University Press}, \bibinfo{year}{1991}).

\bibitem[{\citenamefont{Koshizuka and Ushioda}(1980)}]{KOSHIZUKA:1980sj}
\bibinfo{author}{\bibfnamefont{N.}~\bibnamefont{Koshizuka}} \bibnamefont{and}
  \bibinfo{author}{\bibfnamefont{S.}~\bibnamefont{Ushioda}},
  \bibinfo{journal}{Physical Review B} \textbf{\bibinfo{volume}{22}},
  \bibinfo{pages}{5394} (\bibinfo{year}{1980}).

\end{thebibliography}
\end{document}